\documentclass[journal, doublecolumn, twosided]{IEEEtran}

\usepackage{cite}
\usepackage{url}
\usepackage{float}
\usepackage{verbatim}
\usepackage{graphicx} 
\usepackage{amsmath}
\usepackage{amssymb}
\usepackage{algorithm} 
\usepackage{algorithmic}
\usepackage{color}
\begin{document}
\title{A Neural Network Prediction Based Adaptive Mode Selection Scheme in Full-Duplex Cognitive Networks}
\author{Yirun~Zhang,
        Qirui~Wu,
        Jiancao~Hou,
        Vahid Towhidlou,
        and~Mohammad~Shikh-Bahaei

\thanks{This is a shortened version for conference submission. The full journal version will be available soon on IEEE Transactions on Cognitive Communications and Networking.}
\thanks{The authors are with Center of Telecommunications Research, King's College London, London, United Kingdom, WC2R 2LS. E-mail: \{yirun.zhang@kcl.ac.uk, qirui.wu@kcl.ac.uk, jiancao.hou@kcl.ac.uk, vahid.towhidlou@kcl.ac.uk, m.sbahaei@kcl.ac.uk\}}
}
\maketitle
\IEEEpeerreviewmaketitle

\begin{abstract}
We propose a neural network (NN) predictor and an adaptive mode selection scheme for the purpose of both improving secondary user's (SU's) throughput and reducing collision probability to the primary user (PU) in full-duplex (FD) cognitive networks. SUs can adaptively switch between FD transmission-and-reception (TR) and transmission-and-sensing (TS) modes based on the NN prediction results for each transmission duration. The prediction performance is then analysed in terms of prediction error probability. We also compare the performance of our proposed scheme with conventional TR and TS modes in terms of SU’s average throughput and collision probability, respectively. Simulation results show that our proposed scheme achieves even better SU’s average throughput compared with TR mode. Meanwhile, the collision probability can be reduced close to the level of TS mode.
\end{abstract}

\section{Introduction}

\IEEEPARstart{S}{pectrum} resource scarcity motivates researchers to develop new spectrum management techniques \cite{FD,MSB1,CRnew1,CRnew2}. Cognitive radio networks (CRNs), in which secondary users (SUs) are allowed to sense and access the spectrum white spaces by using dynamic spectrum access (DSA) technology, has become a promising approach to further improve SU's spectrum efficiency \cite{DSAnew,CRnew1}. On the other hand, full-duplex (FD) communications, which allow devices to transmit and receive over the same frequency band at the same time, has also emerged as a solution to solve the spectrum scarcity problem \cite{FD,FDCR,SecReviewer3}. FD operation brings two new transmission modes to CRNs, which are called transmission-and-reception (TR) and transmission-and-sensing (TS) mode, respectively \cite{FDCR,exp,AMC1,Teddy,daoshi2}. In TR mode, SU can achieve higher throughput due to the simultaneous transmission and reception operation at the cost of lower spectrum awareness capability. In TS mode, the collision probability is reduced because of the continuous sensing operation at the expense of lower throughput. These two modes give rise to a trade-off between higher SU's throughput and lower collision probability. Therefore, it is necessary for SU to adaptively switch between these two transmission modes based on different PU's activity in order to optimise the trade-off.

In recent years, researchers have paid attention to the problem of mode selection between FD and half-duplex. The authors in \cite{TSRA} propose an adaptive mode selection scheme based on SU's belief regarding the idleness of the PU. The belief update policy is partially based on calculating the probability of PU's return. Such method requires \textit{a priori} knowledge of PU's activity and signal pattern, which is not practical in the case that PU is not willing to share these information.

In practice, the statistics of PU are usually unknown to SUs, which need to be estimated accordingly. In order to learn and predict PU's dynamics, spectrum prediction methods for CRNs have been studied using different machine learning techniques, such as hidden Markov models (HMMs) \cite{HMM1} and neural network (NN) \cite{NNP}. Among them, the HMM-based prediction methods such as the one used in \cite{HMM1} have several drawbacks, for example, determining an optimal number of states in the HMM is difficult, and the computationally expensive online-training process. The authors in \cite{NNP} propose an NN-based spectrum occupancy prediction method using off-line training, which requires less memory space and lower computational complexity during implementation. All of these existing spectrum prediction works only aim at deciding whether the very next time slot is occupied by PU or not. This one time slot, however, contains limited information on PU's future activity, which may be outdated \cite{HMM3}. Therefore, we design an NN predictor to predict multiple time slots for transmission mode selection in this paper. Based on PU's future activity in multiple time slots, SUs can adaptively switch between different transmission modes either to improve its throughput and or to avoid the collision to PU.

The contributions of this paper are as follows. First, we design a multi-layer NN predictor to solve the channel status prediction problem. The NN predictor we designed can effectively classify the future channel occupancy status into two classes, one refers to the totally unoccupied status and the other represents that some slots will be occupied by PU. Second, we propose a neural network based adaptive mode selection (NN-AMS) scheme by which the SU can utilise the spectrum white spaces efficiently while avoiding the collision to PU. Our proposed NN-AMS scheme is applicable to any random distribution of PU's activity pattern. Third, we define and test the prediction error probabilities. Fourth, we derive SU's average throughput and collision probability of our proposed NN-AMS scheme. We find the optimal length of transmission duration for both NN-AMS scheme and TR mode via simulation of the NN and numerical analysis. Under the optimal length of transmission duration, we address the conclusion that in low sensing error probability scenario, our proposed NN-AMS scheme shows better performance than the conventional TR and TS modes in terms of collision probability and SU's average throughput, respectively.

The rest of this paper is organised as follows. In Section II, we explain our proposed NN-AMS scheme in detail. The performance of the NN-AMS scheme in terms of prediction error probability, SU's average throughput and collision probability are analysed in Section III. Simulation results are then presented in Section IV. Finally, we draw the conclusion in Section V.

\section{The Proposed NN-AMS Scheme}
\begin{figure}[tb] 
\centering
\includegraphics[width=2.5in]{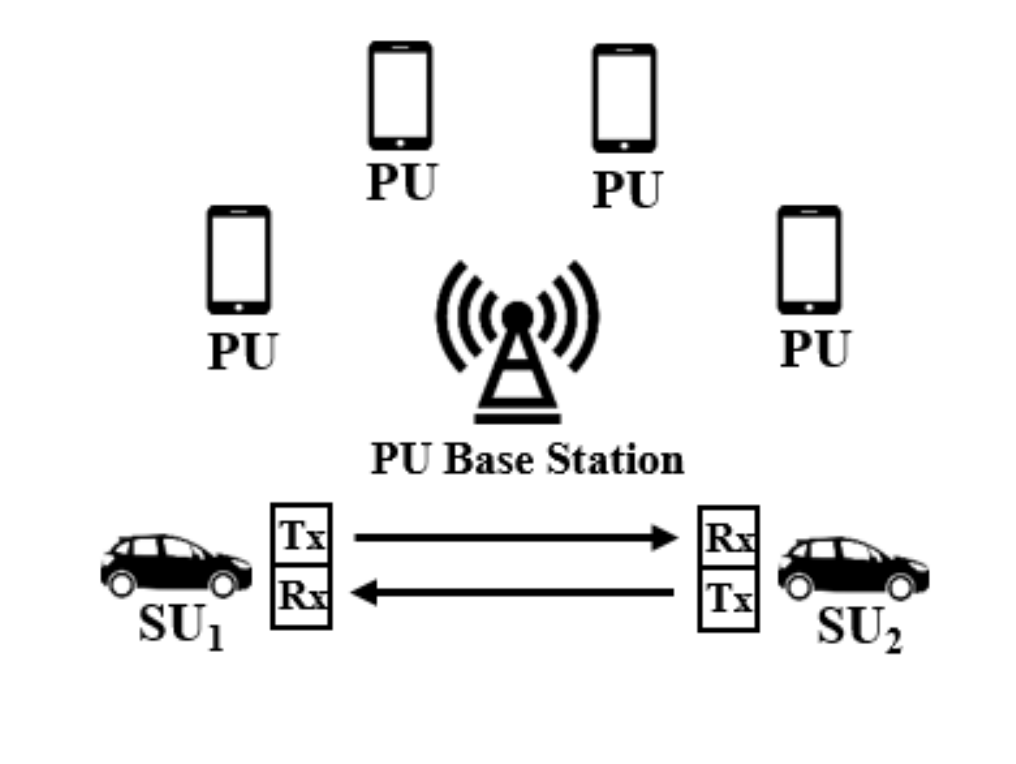}
\caption{System model.}
\label{Figure_1} 
\end{figure}

As shown in Fig. \ref{Figure_1}, we consider a CRN consisting of a PU base station with multiple licensed PUs and a pair of unlicensed SUs which opportunistically access the PU-licensed channel. SUs are equipped with FD radios with partial SIS capability. For fairness and generality, we assume that each SU is equipped with two antennas $\rm Ant_{1}$ and $\rm Ant_{2}$, where $\rm Ant_{1}$ is used for data transmission and $\rm Ant_{2}$ is used for reception. It is worth noting that when operating in TS mode, the transmitter SU (e.g., $\rm SU_{1}$) uses both $\rm Ant_{1}$ and $\rm Ant_{2}$ for simultaneous data transmission and sensing, whereas the receiver SU (e.g., $\rm SU_{2}$) uses only $\rm Ant_{2}$ for receiving data from the transmitter SU.

PU's traffic is modelled as a simple ON and OFF process with average lengths of \(T_{0}\) and \(T_{1}\) respectively, based on whether the channel is occupied by PU or not \cite{ONOFF}. The activity pattern of PU's traffic follows arbitrary random distribution whose parameters are unknown to the SUs.

\begin{figure*}[htb] 
\centering
\includegraphics[width=6in]{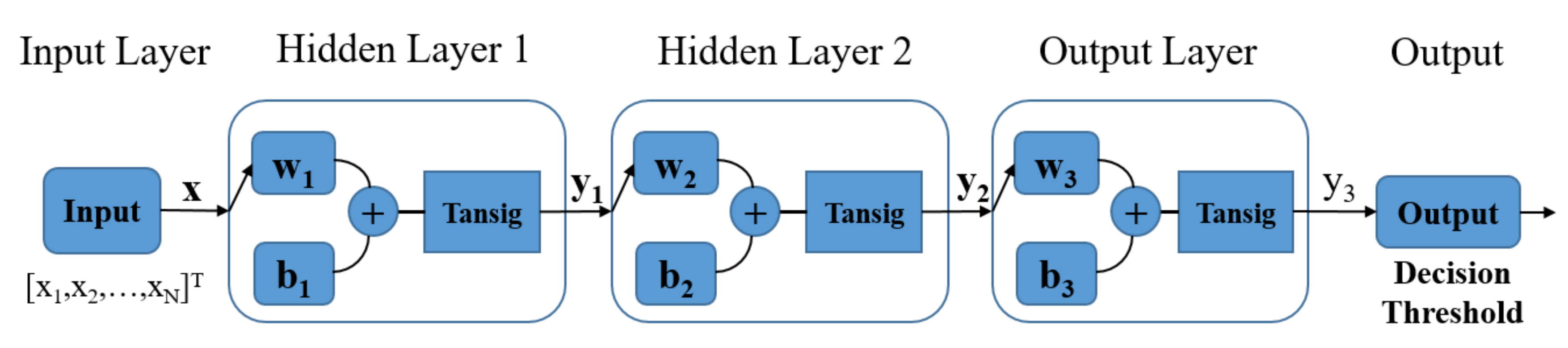} 
\caption{The structure of the neural network predictor.}
\label{Figure_2} 
\end{figure*} 

\subsection{Energy-Detection Based Sensing}
The energy-detection based spectrum sensing technique is applied to the NN prediction due to its simplicity and ability to identify spectrum white spaces without requiring any \textit{a priori} information of PU's signal pattern \cite{SecReviewer1,ED}. The sensing result contains two states: ``1'' is for busy; ``-1'' is for idle. The performance of spectrum sensing is judged by two fundamental measures, the detection probability (\(P_{d}\)) and the false-alarm (\(P_{f}\)) probability. The former refers to the probability if PU is occupying the channel, SU can successfully detect them. The latter is the probability if the channel is idle, SU falsely decides that it is occupied by PU. 

We have to consider the residual SI when deriving the false-alarm and detection probabilities. Here we do not elaborate on detail of the applied SIS methods and just quantify this capability by $\chi$, where \(\chi=0\) corresponds to perfect SIS (i.e., no residual SI) and \(\chi=1\) corresponds to the case where the SI is not suppressed.

From \cite{exp}, the two probabilities considering SI (e.g., the continuous sensing operations in TS mode) are given by:\vspace{-0.5em}
\begin{equation}
\begin{aligned}
P_{d,SI}=&Q\left ( \left ( \frac{\varepsilon _{th}}{\sigma ^{2}}-\chi^{2}\rm {SNR_{s,s}}-\rm {SNR_{s,p}}-1\right ) \times \right.\\
&\left. \sqrt{\frac{\omega _{s}T_{s}}{2\chi^{2} \rm {SNR_{s,s}}+2\chi^{2} \rm {SNR_{s,s}SNR_{s,p}}+2\rm {SNR_{s,p}}+1 }} \right )
\label{chi}
\end{aligned}
\end{equation}
\begin{equation}
P_{f,SI}=Q\left ( \left ( \frac{\varepsilon_{th}}{\sigma ^{2}}-\chi^{2}\rm {SNR_{s,s}}-1\right )\sqrt{\frac{\omega _{s}T_{s}}{2\chi^{2}\rm {SNR_{s,s}}+1}} \right ),
\end{equation}
where \(Q(x)=\frac{1}{\sqrt{2\pi}}\int_{x}^{\infty}e^{(-\frac{t^{2}}{2})}dt\) is the \(Q\)-function; \(\varepsilon_{th}\) is the threshold of energy detection; \(\sigma^{2}\) is the variance of the circular symmetric complex Gaussian background noise; \(\omega_{s}\) is the sampling frequency. Moreover, \(\rm {SNR}_{i,j}\) is the signal-to-noise ratio (SNR) transmitted by transmitter \(j\) and received at receiver \(i\). \(i\) and \(j\) can be either \(p\) or \(s\), where \textit{p} and \textit{s} refer to PU and SU respectively. If \(\chi\) equals to zero, the false-alarm probability and detection probability converge to those without SI (e.g., the initial sensing operation of TR mode) which are defined as \(P_{f}\) and \(P_{d}\), respectively.

\subsection{Neural Network Based Prediction}
\subsubsection{Structure Design}
The main idea of our proposed NN predictor is to insert \textit{N} number of past sensing results into the NN predictor and get the estimated PU's future activity of \textit{M} time slots. We introduce one parameter \(m\) to classify PU's future activity of \(M\) time slots. \(m\) refers to the number of idle time slots in the future \(M\) time slots. If \(m \geqslant M\), the output of the NN predictor is ``-1''; Otherwise, its output is ``1''.

The NN predictor contains three kinds of layers, including the input layer, the hidden layer, and the output layer, which is illustrated in Fig. \ref{Figure_2}. The output layer has one neuron. Each layer is fully-connected with adaptive weights, biases and an activation function. We use the Tan-Sigmoid function as the activation function for each layer, which is given by:\vspace{-0.5em}
\begin{equation}
{\rm Tan\textit{-}Sigmoid}:\quad{\rm tansig}(x)=\frac{2}{1+e^{-2x}}-1.
\end{equation}

We define the input vector as \(\textbf{x}=[x_{1},x_{2},...,x_{N}]^{T}\) whose elements are either ``-1'' (idle) or ``1'' (busy). The hidden layers and output layer contain multiple neurons. Each neuron in the network has a weight vector \(\textbf{w}\) and a bias \(b\), whose elements are all real numbers. Then we define the weight vector for neuron \(i\) in hidden layer one as \(\textbf{w}_{1,i}=[w_{1,i}^{1},w_{1,i}^{2},...,w_{1,i}^{N}]^{T}\) which contains \(N\) elements, for neuron \(j\) in hidden layer two as \(\textbf{w}_{2,j}=[w_{2,j}^{1},w_{2,j}^{2},...,w_{2,j}^{15}]^{T}\) which contains 15 elements, and for the neuron in output layer as \(\textbf{w}_{3}=[w_{3}^{1},w_{3}^{2},...,w_{3}^{20}]^{T}\) which contains 20 elements. The biases for neurons in each layer are defined as \(\textbf{b}_{1}=[b_{1}^{1},b_{1,i}^{2},...,b_{1}^{15}]^{T}\), \(\textbf{b}_{2}=[b_{2}^{1},b_{2}^{2},...,b_{2}^{20}]^{T}\) and \(\textbf{b}_{3}=[b_{3}]\), receptively. The output vector for hidden layers are \(\textbf{y}_{1}=[y_{1}^{1},y_{1}^{2},...,y_{1}^{15}]^{T}\) and \(\textbf{y}_{2}=[y_{2}^{1},y_{2}^{2},...,y_{2}^{20}]^{T}\), respectively. The output of the output layer is \(y_{3}\in \{-1,1\}\). The three outputs can be expressed by:\vspace{-0.5em}
\begin{subequations}
\begin{equation}
y_{1}^{i}={\rm tansig}(\sum_{k=1}^{N}w_{1,i}^{k}x_{k}+b_{1}^{i})={\rm tansig}(\textbf{w}_{1,i}\textbf{x}+b_{1}^{i})
\end{equation}\vspace{-0.5em}
\begin{equation}
y_{2}^{j}={\rm tansig}(\sum_{k=1}^{15}w_{2,j}^{k}y_{1}^{k}+b_{2}^{j})={\rm tansig}(\textbf{w}_{2,j}\textbf{y}_{1}+b_{2}^{j})
\end{equation}
\begin{equation}
y_{3}={\rm tansig}(\sum_{k=1}^{20}w_{3}^{k}y_{2}^{k}+b_{3})={\rm tansig}(\textbf{w}_{3}\textbf{y}_{2}+b_{3}),
\end{equation}
\end{subequations}
where \(i \in [1,15]\) and \(j \in [1,20]\). By using a decision threshold at the output, the predicted value \(y_{3}\) can be expressed as a binary symbol:\vspace{-0.5em}
\begin{equation}
\begin{aligned}
& {\rm if}\;y_{3} < 0\quad {\rm then}\;y_{3}\;=\;-1 \\ 
& {\rm if}\;y_{3} \geqslant 0\quad {\rm then}\;y_{3}\;=\;1.
\end{aligned}
\end{equation}

\subsubsection{Preprocessing and Training}
The NN predictor is trained by using \(N_{tr}\) independent training sets. Each training set contains a pair of input vector and output value. First, SU senses the channel for a considerably long period of time and generates a sensing sequence. Then it randomly picks \(N_{tr}\) sets each with length of \(N+M\) from the sensing sequence. Subsequently, SU puts the first \(N\) sensing results into the input vector and counts the number of idle slots in the last \(M\) slots. If \(m \geqslant M\), the SU sets the target output value to ``-1''; Otherwise, it sets the target output value to ``1''. 

After all training sets are preprocessed, SU inputs them into the NN predictor for off-line training. The main idea of NN training is to update the weight vectors \(\textbf{w}\) and biases \(b\) iteratively in order to minimise the mean squared error between the predicted output and the target output. We use the Levenberg-Marquardt Backpropagation (LMBP) algorithm in \cite{LM1} for off-line training due to its quick convergence.

\subsection{Adaptive Mode Selection Scheme}

SU adaptively switches between FD-TR and FD-TS modes, based on the output of the NN predictor. In this section, we first introduce these two transmission modes and then describe the adaptive mode selection algorithm.

\subsubsection{TR mode}
In TR mode, the transmission duration is defined as \(T_{p}=MT_{s}\), where \(T_{s}\) refers to the length of one time slot. SU first senses the primary channel for one time slot with time \(T_{s}\) and then transmits and receives data in FD manner for \(M-1\) time slots with time \(T_{t}=T_{p}-T_{s}\) which is the data transmission duration. If the initial sensing gives a busy result, then the SU switches to the TS mode instead of transmitting and receiving. SU can improve its throughput in TR mode, but may cause a collision to PU since SU cannot be aware of the return of PU during its data transmission and reception. 

\subsubsection{TS mode}
In TS mode, after the initial sensing period \(T_{s}\), SU senses the channel continuously during data transmission. The duration of data transmission is divided into multiple time slots with the same length of time as the initial sensing period. The SU transmits data whenever it senses the channel to be idle. If at the end of any time slot the channel status is detected to be busy, SU immediately stops its transmission. Then SU can alleviate collision to PU in TS mode at the expense of lower throughput than in TR mode because it cannot simultaneously transmit and receive data while sensing.

For adaptive switching between TR and TS modes, the determination of a proper switching threshold is critical. Based on this principle, SU chooses TR mode in order to improve the spectrum efficiency if the output is ``-1''; Otherwise, it chooses to transmit in TS mode in order to avoid collision with PU. 

The input vector is updated at the end of each transmission duration. Depending on which transmission mode SU has chosen, different update policy is applied. If SU has chosen TR mode and the initial sensing result is ``-1'' (i.e., SU starts to transmit data in TR mode), SU adds the prediction results (i.e., \(M\) zeros) to the end of the input vector and deletes the oldest \(M\) elements. If SU has chosen TR mode and the initial sensing result is ``1'', or it has chosen TS mode, it simply updates the input vector using \(M\) sensing results. The NN-AMS scheme follows the general steps described in Algorithm \ref{Algorithm1}.

\begin{algorithm}[tb]
\caption{Proposed Adaptive Mode Selection Scheme}
\label{Algorithm1}
\textbf{Begin While} (\textit{Next M Time Slots Prediction} is \textbf{NOT NULL}) \\
\textbf{Step-1}: \textit{Input Vector Initialisation}: Initialise the input vector \textbf{x} by sensing \textit{N} time slots. \\
\textbf{Step-2}: \textit{Spectrum Occupancy Prediction}: Input \textbf{x} into the NN predictor and get the output \(y_{3}\). \\
\textbf{Step-3}: \textit{Mode Selection Decision}:

\qquad\quad\ \textbf{If} \(y_{3}=-1\) \textbf{And} \(\rm {initial\;sensing}=-1\) \textbf{Then}

\qquad\quad\quad\ SU transmits in TR mode in M slots.

\qquad\quad\ \textbf{Else}

\qquad\quad\quad\ SU transmits in TS mode in M slots.

\qquad\quad\ \textbf{End IF} \\
\textbf{Step-4}: \textit{Input Vector Update}:

\qquad\quad\ \textbf{If} \(y_{3}=-1\) \textbf{And} \(\rm {initial\;sensing}=-1\) \textbf{Then}

\qquad\quad\quad\ Use the prediction result as new sensing results.

\qquad\quad\ \textbf{Else}

\qquad\quad\quad\ Use new sensing results.

\qquad\quad\ \textbf{End IF}

\qquad\quad\ Delete the first \(M\) elements from the input vector \textbf{x} and add the new sensing results to the end, and go to \textbf{Step-2}. \\
\textbf{End While}
\end{algorithm}

\section{Performance Analysis}
In this section, we first define the prediction error probabilities. Then we derive SU's average throughput and collision probability of NN-AMS scheme.

\subsection{Prediction Error Probabilities}
The accuracy of prediction can be judged by testing with \(N_{tt}\) independent testing sets and computing the prediction output. Each testing set is generated by randomly observing \(N\) time slot and predicted the spectrum occupancy status of future \(M\) time slots. We compare the prediction results of testing with true spectrum occupancy status and calculate the prediction false-alarm probability \(P_{pf}\), the prediction detection probability \(P_{pd}\), and the average prediction error probability \(P_{e}\). If the prediction result is different from the true spectrum occupancy status, we call it an incorrect prediction.

The prediction false-alarm probability \(P_{pf}\) refers to the probability that the prediction result shows ``-1'' while the true result is ``1'', which is defined as:\vspace{-0.5em}
\begin{equation}
P_{pf}={\rm Pr}[y_{3}=1|m=M]=\frac{N_{01}}{N_{0}}, \label{ppf}
\end{equation}
where \(N_{01}\) refers to the number of sets that the SU falsely detects the channel will be occupied by the PU when there are no PU activities in the next \(M\) slots; \(N_{0}\) is the number of sets that there are no PU activities in the next \(M\) slots among \(N_{tt}\) testing sets.

The prediction detection probability \(P_{pd}\) refers to the probability that the prediction result shows ``-1'' while the true result is also ``-1'', which is given by:\vspace{-0.5em}
\begin{equation}
P_{pd}={\rm Pr}[y_{3}=1|m<M]=1-\frac{N_{e}-N_{01}}{N_{tt}-N_{0}}, \label{ppd}
\end{equation}
where \(N_{e}\) refers to the total number of incorrect prediction sets. 

The average prediction error probability \(P_{e}\), which refers to the probability that the prediction result is different from the true result, is defined as:\vspace{-0.5em}
\begin{equation}
P_{e}=\frac{N_{e}}{N_{tt}}.
\end{equation}

\subsection{Secondary User's Average Throughput}
\begin{figure}[tb] 
\centering
\includegraphics[width=3.5in]{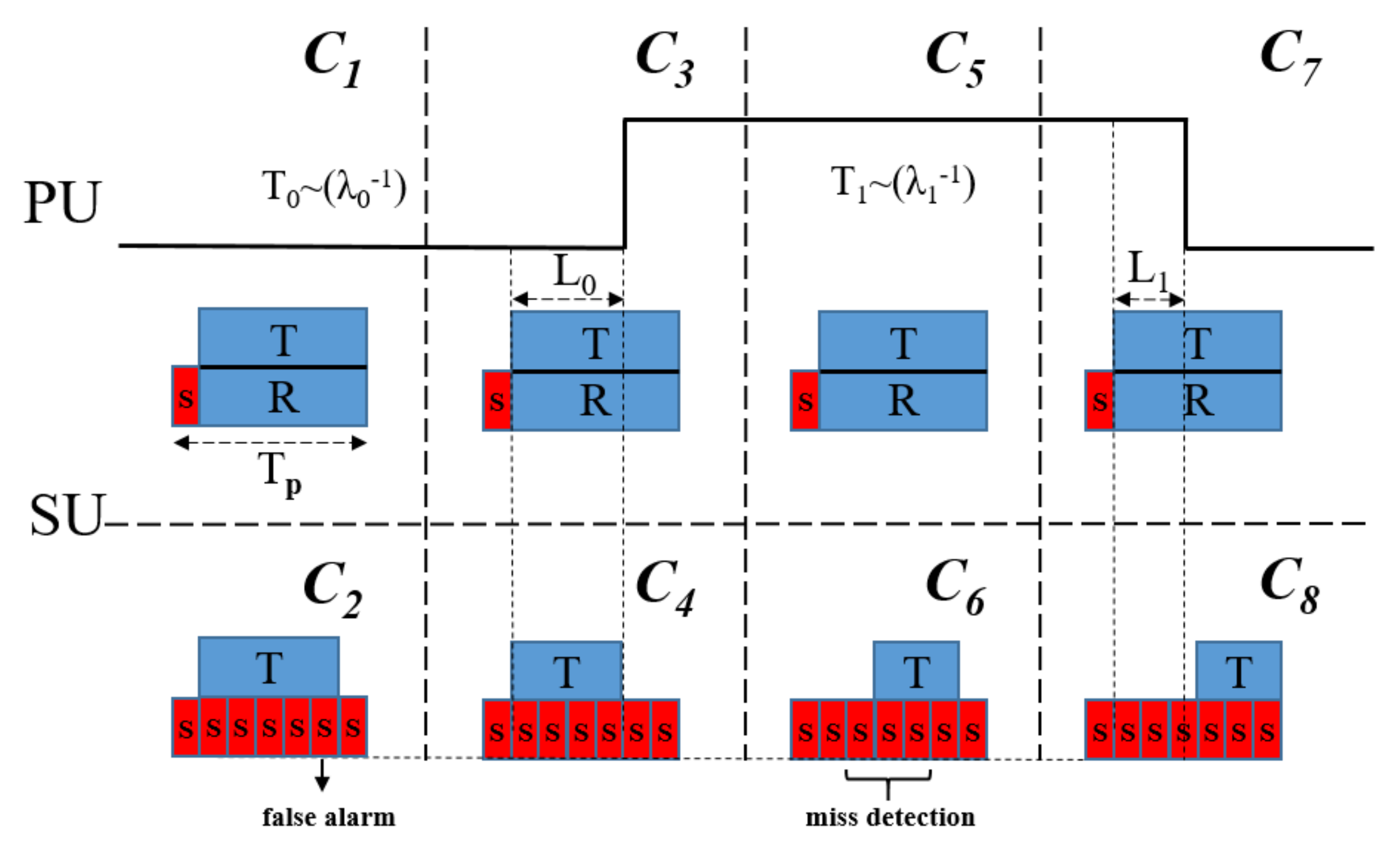} 
\caption{Eight different cases of SU's transmission in the NN-AMS scheme.}
\label{Figure_3} 
\end{figure} 

In this section, we derive the SU's average throughput for NN-AMS scheme, TR and TS mode. The average throughput contains throughputs in different cases, which can be defined as:\vspace{-0.5em}
\begin{equation}
R_{avg}=\sum_{i=1}^{n}{\rm Pr}[C_{i}]R_{i},
\end{equation}
where \(n\) is the number of cases; \(C_{i}\) refers to the \(i\)th case; \({\rm Pr}[C_{i}]\) is occurrence probability of the \(i\)th case; \(R_{i}\) is SU's throughput in the \(i\)th case.

Our proposed NN-AMS scheme is applicable under arbitrary PU's activity distribution. However, in order to fairly compare with the existing works \cite{exp,Teddy,vahid}, we assume the length of the ON and OFF state follows the negative exponential distribution with parameter \(\lambda_{1}\) and \(\lambda_{0}\), respectively. Note that SU's transmission duration \(T_{p}\) is considered small relative to \(T_{1}\) and \(T_{0}\) (\(T_{p}\ll T_{0}\;{\rm and}\;T_{1}\)). Under this assumption, the scenario where one \(T_{p}\) contains more than one ON and OFF transition rarely happens\cite{exp}. 

In our proposed NN-AMS scheme, there are eight different cases of SU's transmission, \(C_{1}\) to \(C_{8}\), according to different beginnings and endings of SU's one transmission duration and different prediction results, as shown in Fig. \ref{Figure_3}. Since the last four cases are symmetrical to the first four, we only derive the first four cases here for simplicity. \footnote{Full derivation will be available in our journal paper.}

\begin{itemize}
\item \(C_{1}\)
\end{itemize}

In \(C_{1}\), the total transmission duration falls into the OFF state.  Since the initial sensing operation in TR mode is not influenced by SI, sensing error probabilities \(P_{f}\) and \(P_{d}\) are used instead of \(P_{f,SI}\) and \(P_{d,SI}\). SU decides to transmit in TR mode and senses the channel to be idle with probability \(\overline{P}_{pf} \cdot \overline{P}_{f}\), where \(\overline{P}_{pf}=1-P_{pf}\) is the complementary probability of prediction false-alarm probability \(P_{pf}\); \(\overline{P}_{f}=1-P_{f}\) is the complementary probability of false-alarm probability \(P_{f}\). We define the forward recurrence time of the OFF state as \(L_{0}\), which refers to the length from the start of SU's transmission to the end of the OFF state. According to the memoryless property of exponential distribution, \(L_{0}\) also follows exponential distribution with parameter \(\lambda_{0}\). The occurrence probability of this case is given by:\vspace{-0.5em}
\begin{equation}
{\rm Pr}[C_{1}]=\mu_{0} \cdot \overline{P}_{pf} \cdot \overline{P}_{f} \cdot e^{-\frac{T_t}{\lambda_{0}}},
\label{C1}
\end{equation}
where \(\mu_{0}=\frac{\lambda_{0}}{\lambda_{0}+\lambda_{1}}\) refers to the occurrence probability of OFF state. Due to the simultaneous transmission-and-reception, the achievable throughput should be twice as high as that in half-duplex scheme \cite{Teddy}. Therefore, SU's average throughput \(R_{1}\) of this case can be written as:\vspace{-0.5em}
\begin{equation}
R_{1}=2 \cdot \frac{T_t}{T_p} \cdot D_{0,TR},
\label{R1}
\end{equation}
where \(D_{0,TR}={\rm log}_{2}(1+\frac{{\rm SNR_{s,s}}}{1+\chi {\rm SNR_{s,s}}})\) is SU's maximum achievable data rate in TR mode in the OFF state \cite{vahid}.

\begin{itemize}
\item \(C_{2}\)
\end{itemize}

This case includes two situations. First, SU makes an incorrect prediction with probability \(P_{pf}\) and decides to transmit in TS mode. Second, the SU decides to transmit in TR mode initially with probability \(\overline{P}_{pf}\); However, it switches to TS mode after it falsely senses the channel to be busy with probability \(P_{f}\). The occurrence probability of this case is given by:\vspace{-0.5em}
\begin{equation}
{\rm Pr}[C_{2}]=\mu_{0} \cdot (1-\overline{P}_{pf} \cdot \overline{P}_{f}) \cdot e^{-\frac{T_t}{\lambda_{0}}}.
\label{C2}
\end{equation}

In a conservative manner, the influence of SI on sensing in TS mode is considered over all time slots due to simultaneous sensing-and-transmission. Since the sensing operation in each time slot is independent, the event where SU transmits data follows binomial distribution with probability of occurrence given by \(\overline{P}_{f,SI}\), where \(\overline{P}_{f,SI}=1-P_{f,SI}\) is the complementary probability of false-alarm probability \(P_{f,SI}\). Thus, SU's average throughput in this case is given by:\vspace{-0.5em}
\begin{equation}
R_{2}=\sum_{j=1}^{M-1}\binom {M-1}{j}\overline{P}_{f,SI}^{j}P_{f,SI}^{(M-1-j)} \frac{j}{M} \cdot D_{0,TS},
\label{R2}
\end{equation}
where \(j\) refers to the number of time slots that the SU transmits data when the channel is idle; \(D_{0,TS}={\rm log}_{2}(1+{\rm SNR_{s,s}})\) is SU's maximum achievable data rate in TS mode in the OFF state.

\begin{itemize}
\item \(C_{3}\)
\end{itemize}

This case indicates that PU comes back during SU's data transmission. SU decides to transmit in TR mode with probability \(\overline{P}_{pd}\) and the initial sensing period gives an idle result with probability \(\overline{P}_{f}\), where \(\overline{P}_{pd}=1-P_{pd}\) is the complementary probability of prediction detection probability \(P_{pd}\). In this case, \(L_{0}\) is larger than zero but smaller than \(T_t\). Thus, the occurrence probability of this case is given by:\vspace{-0.5em}
\begin{equation}
{\rm Pr}[C_{3}]=\mu_{0} \cdot \overline{P}_{pd} \cdot \overline{P}_{f} \cdot (1-e^{-\frac{T_t}{\lambda_{0}}}).
\label{C3}
\end{equation}

SU's average throughput of this case contains two parts, one without collision to PU and the other with collision to PU. The instantaneous average throughput of SU of this case is given by:\vspace{-0.5em}
\begin{equation}
R_{3}(\hat{L}_{0})=2 \cdot \left(\frac{D_{0,TR}-D_{1,TR}}{T_p} \cdot \hat{L}_{0}+\frac{T_t}{T_p} \cdot D_{1,TR}\right),
\label{R3}
\end{equation}
where \(D_{1,TR}={\rm log}_{2}(1+\frac{{\rm SNR_{s,s}}}{1+{\rm SNR_{s,p}}+\chi {\rm SNR_{s,s}}})\) is SU's maximum achievable data rate in TR mode in the ON state \cite{vahid}; \(\hat{L}_{0}\) is the normalised recurrence time of the OFF state, whose probability density function (p.d.f) is given by:\vspace{-0.5em}
\begin{equation}
\hat{l}_{0}= \frac{\frac{1}{\lambda_{0}}e^{-\frac{t}{\lambda_{0}}}}{1-e^{-\frac{T_t}{\lambda_{0}}}}.
\end{equation}

Since \(\hat{L}_{0}\) is a random variable, eq. (\ref{R3}) is the SU's throughput with a determined \(\hat{L}_{0}\). Its expected value is expressed as:\vspace{-0.5em}
\begin{equation}
\begin{aligned}
\mathbb{E}(R_{3})=&2 \cdot \frac{D_{0,TR}-D_{1,TR}}{T_p} \cdot \frac{\lambda_{0}-\left ( T_t +\lambda_{0}\right )e^{-\frac{T_t}{\lambda_{0}}}}{1-e^{-\frac{T_t}{\lambda_{0}}}} \\
&+ 2 \cdot \frac{T_t}{T_p} \cdot D_{1,TR}.
\end{aligned}
\end{equation}

\begin{itemize}
\item \(C_{4}\)
\end{itemize}

Case \(C_{4}\) happens when SU makes a correct prediction that PU will come back in the next \(M\) time slots and transmit in TS mode with probability \(P_{pd}\), or it decides to transmit in TR mode but the initial sensing period gives a busy result with probability \(\overline{P}_{pd} \cdot P_{f}\). The occurrence probability of this case is given by:\vspace{-0.5em}
\begin{equation}
{\rm Pr}[C_{4}]=\mu_{0} \cdot (1-\overline{P}_{pd} \cdot \overline{P}_{f}) \cdot (1-e^{-\frac{T_t}{\lambda_{0}}}).
\label{C4}
\end{equation}

In the analysis of SU's average throughput of TS mode, if PU comes back during a time slot, we assume that this certain time slot is in the ON state. Since the difference of throughput in one single time slot is extremely small, we can use discrete function to express SU's average throughput. SU's average throughput of this case is more complicated than that of \(C_{2}\) since SU can transmit \(j\) (\(1 \leqslant j \leqslant \lfloor\frac{\hat{L}_{0}}{T_{s}}\rfloor\)) time slots in the OFF state and \(k\) (\(1 \leqslant k \leqslant M-1-\lfloor\frac{\hat{L}_{0}}{T_{s}}\rfloor\)) time slots in the ON state. \(\lfloor.\rfloor\) is the floor function that takes as input a real number \(x\) and gives as output the greatest integer less than or equal to \(x\). Due to the floor function, it is complicated to find the expected value of \(R_{4}\). Therefore, we use \(\mathbb{E}(\hat{L}_{0})\) to approximate the expected value of \(R_{4}\) expressed as:\vspace{-0.5em}
\begin{equation}
\begin{aligned}
\mathbb{E}(R_{4}) \approx &\sum_{j=1}^{\lfloor \frac{\mathbb{E}(\hat{L}_{0})}{T_{s}}\rfloor}\binom {\lfloor \frac{\mathbb{E}(\hat{L}_{0})}{T_{s}}\rfloor}{j}\overline{P}_{f,SI}^{j}P_{f,SI}^{\lfloor \frac{\mathbb{E}(\hat{L}_{0})}{T_{s}}\rfloor-j} \frac{j}{M} \cdot D_{0,TS}\\
&+\sum_{k=1}^{M-1-\lfloor \frac{\mathbb{E}(\hat{L}_{0})}{T_{s}}\rfloor}\binom {M-1-\lfloor \frac{\mathbb{E}(\hat{L}_{0})}{T_{s}}\rfloor}{k}\overline{P}_{d,SI}^{k} \\
&\times P_{d,SI}^{M-1-\lfloor \frac{\mathbb{E}(\hat{L}_{0})}{T_{s}}\rfloor-k}\frac{k}{M} \cdot D_{1,TS},
\end{aligned}
\end{equation}
where \(k\) refers to the number of time slots that SU transmits in the ON state; \(\overline{P}_{d,SI}=1-P_{d,SI}\) is the complementary probability of detection probability \(P_{d,SI}\); \(D_{1,TS}={\rm log}_{2}(1+\frac{{\rm SNR_{s,s}}}{1+{\rm SNR_{s,p}}})\) is SU's maximum achievable data rate in TS mode in the ON state.

Finally, SU's average throughput of our proposed NN-AMS scheme taking all cases into account is given by:\vspace{-0.5em}
\begin{equation}
R_{{\rm NN-AMS}}=\sum_{i=1}^{8}{\rm Pr}[C_{i}]R_{i}.
\label{RNNAMS}
\end{equation}

\subsection{Collision probability}
In general, two different events lead to collision. First, PU comes back during the SU's data transmission, for example, \(C_{3}\) in Fig. \ref{Figure_3}. This kind of event occurs when SU transmits in TR mode. Second, SU fails to detect the presence of PU and starts to transmit data as in \(C_{5}\) and \(C_{6}\) in Fig. \ref{Figure_3}. This type of collision occurs both in TR and TS mode, whose occurrence probability depends on the detection probability.

The collision probability of NN-AMS scheme consists of all cases except \(C_{1}\) and \(C_{2}\). If SU chooses to transmit in TR mode, collision probabilities are the same as the occurrence probabilities. Note that if SU chooses to transmit in TS mode, we use the complementary probability of non-collision case (i.e., SU successfully detects all the busy channel state) to calculate the collision probability. Collision probabilities of \(C_{4}\), \(C_{6}\) and \(C_{8}\) are given by:\vspace{-0.5em}
\begin{equation}
\begin{aligned}
{\rm Pr}[C_{4,C}]=\mu_{0} \cdot (1-\overline{P}_{pd} \cdot \overline{P}_{f}) \cdot (1-e^{-\frac{T_t}{\lambda_{0}}})  \overline{P}_{d,SI}^{M-1-\lfloor \frac{\mathbb{E}(\hat{L}_{0})}{T_{s}}\rfloor},
\end{aligned}
\end{equation}\vspace{-0.5em}
\begin{equation}
{\rm Pr}[C_{6,C}]=\mu_{1} \cdot (1-\overline{P}_{pd} \cdot \overline{P}_{d}) \cdot e^{-\frac{T_t}{\lambda_{1}}} \cdot \overline{P}_{d,SI}^{M-1},
\end{equation}\vspace{-0.5em}
\begin{equation}
\begin{aligned}
{\rm Pr}[C_{8,C}]=\mu_{1} \cdot (1-\overline{P}_{pd} \cdot \overline{P}_{d}) \cdot (1-e^{-\frac{T_t}{\lambda_{1}}}) \overline{P}_{d,SI}^{\lceil \frac{\mathbb{E}(\hat{L}_{1})}{T_{s}}\rceil}.
\end{aligned}
\end{equation}

Then the collision probability of NN-AMS scheme considering all cases is given by:\vspace{-0.5em}
\begin{equation}
\begin{aligned}
P_{{\rm NN-AMS,C}}=&{\rm Pr}[C_{3}]+{\rm Pr}[C_{4,C}]+{\rm Pr}[C_{5}] \\
&+ {\rm Pr}[C_{6,C}]+{\rm Pr}[C_{7}]+{\rm Pr}[C_{8,C}].
\end{aligned}
\end{equation}

\begin{figure}[tb] 
\centering
\includegraphics[width=3.5in]{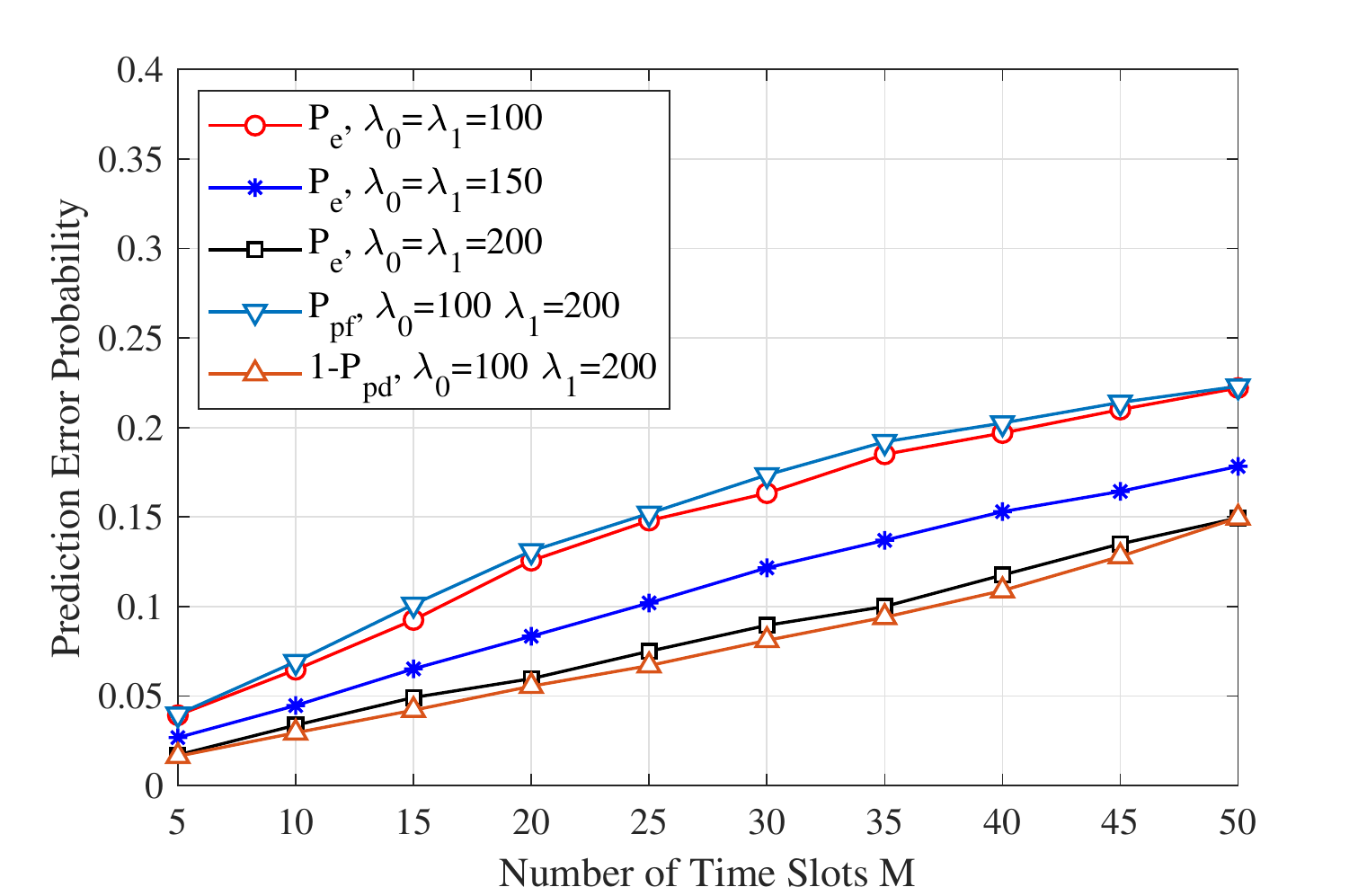} 
\caption{Prediction error probability versus the number of time slots \(M\) (perfect sensing).}
\label{Figure_4} 
\end{figure} 

\section{Simulation Results}

In this section, we demonstrate simulation results for prediction error probabilities, collision probabilities, and average throughput of SU. We set the sampling frequency \(\omega_{s}=6\;\)MHz, SIS coefficient \(\chi=0.1\), \(\rm {SNR}_{s,s}=10\;\)dB, \(\rm {SNR}_{s,p}=9\;\)dB, \(T_{s}=0.001\) sec, and \(\lambda_{0}=\lambda_{1}=0.1\) sec. The number of input $N$, training sets \(N_{tr}\) and testing sets \(N_{tt}\) are set to be 75, 1000 and 30000, respectively.

\subsection{Prediction Error Probability Analysis}
Figure. \ref{Figure_4} illustrates prediction error probability versus the number of time slots predicted \(M\), assuming the sensing is perfect. As shown in Fig. \ref{Figure_4}, all three prediction error probabilities increase with \(M\) due to the increase of randomness. As the average lengths of ON and OFF state increase, the prediction error probability decreases. This indicates that the NN predictor can predict more time slots with the same prediction error probability when \(\lambda_{0}\) and \(\lambda_{1}\) increase. It can also be seen that the prediction false-alarm probability and prediction detection probability are related to \(\lambda_{0}\) and \(\lambda_{1}\), respectively. These two probabilities are close to the average prediction probability of the same \(\lambda_{0}\) and \(\lambda_{1}\). 

\subsection{SU's Average Throughput Analysis}

\begin{figure}[tb] 
\centering
\includegraphics[width=3.5in]{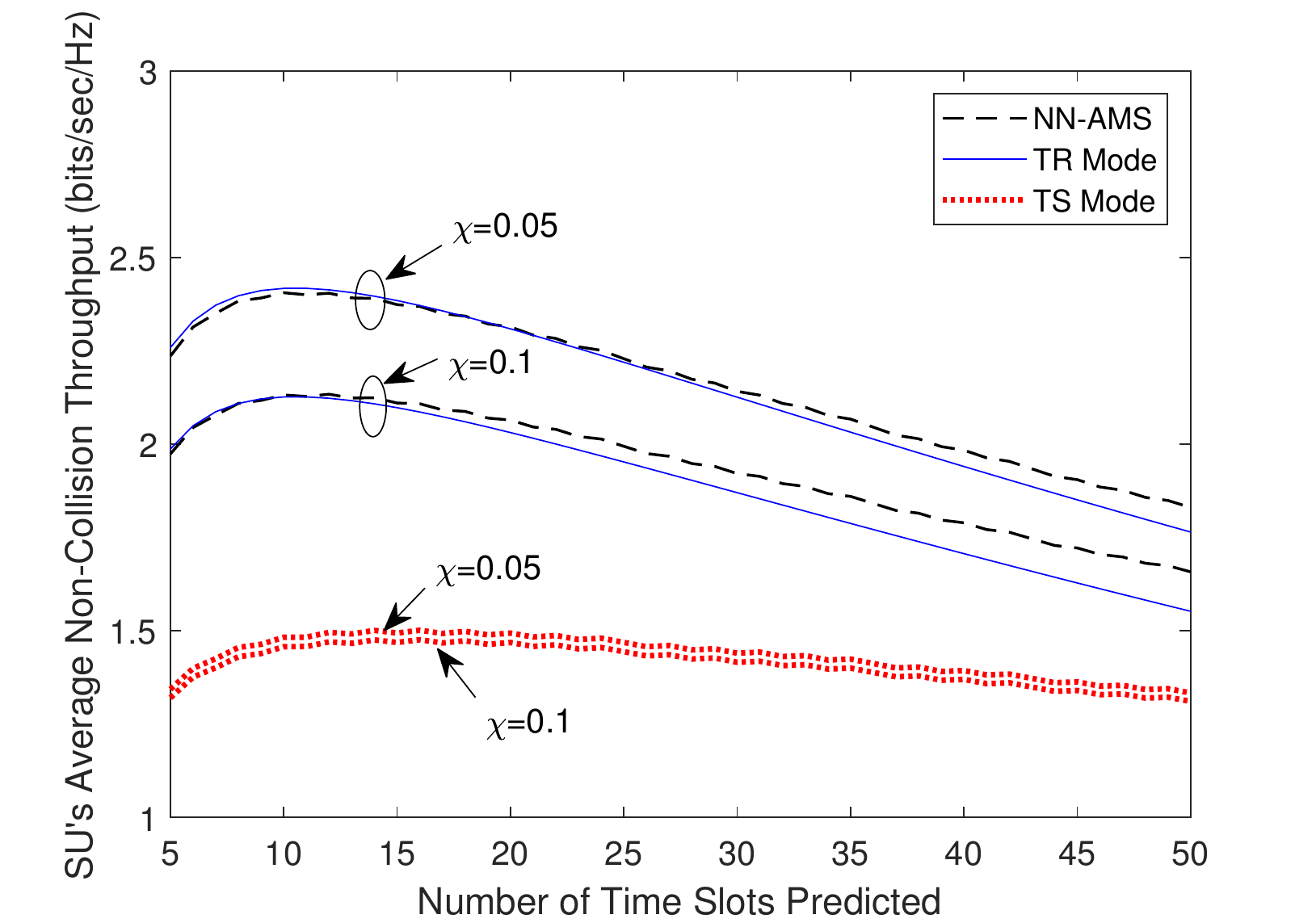} 
\caption{SU's non-collision average throughput versus the number of time slots \(M\).}
\label{Figure_5} 
\end{figure}

Figure. \ref{Figure_5} depicts SU's average throughput versus $M$, assuming that PU has zero tolerance for collision (i.e., the non-collision case). It can be seen that SU's average throughput of TS mode is hardly influenced by $M$ because it only depends on the data rates and sensing error probabilities. SU's average throughput of NN-AMS scheme is very close to that of TR mode with different values of \(M\) and \(\chi\). With large values of $M$, SU's average throughput of NN-AMS scheme is slightly larger than that of TR mode. This indicates that our proposed NN-AMS scheme can successfully detect the return of PU and switch to TS mode in order to avoid collision to PU, and thus, increase the SU's average throughput. On the other hand, we can see that SU's average throughput first increases then decreases after it reaches the maximal value for both NN-AMS scheme and TR mode. From  the figure, we can find the optimal \(M^{*}\) for both NN-AMS scheme and TR mode is 10 in this case. This indicates that SU can achieve its maximum non-collision average throughput when its transmission duration is set to be \(T_{p}=MT_{s}=0.01\) sec if \(\lambda_{0}=\lambda_{1}=0.1\) sec. Note that for difference values of \(\lambda_{0}\) and \(\lambda_{1}\), the optimal value of \(M\) is different. Through jointly analysing Fig. \ref{Figure_4} and Fig. \ref{Figure_5}, we can draw a conclusion that the optimal \(M^{*}\) is larger for larger values of \(\lambda_{0}\) and \(\lambda_{1}\) due to lower prediction error probabilities.

\begin{figure}[tb] 
\centering
\includegraphics[width=3.5in]{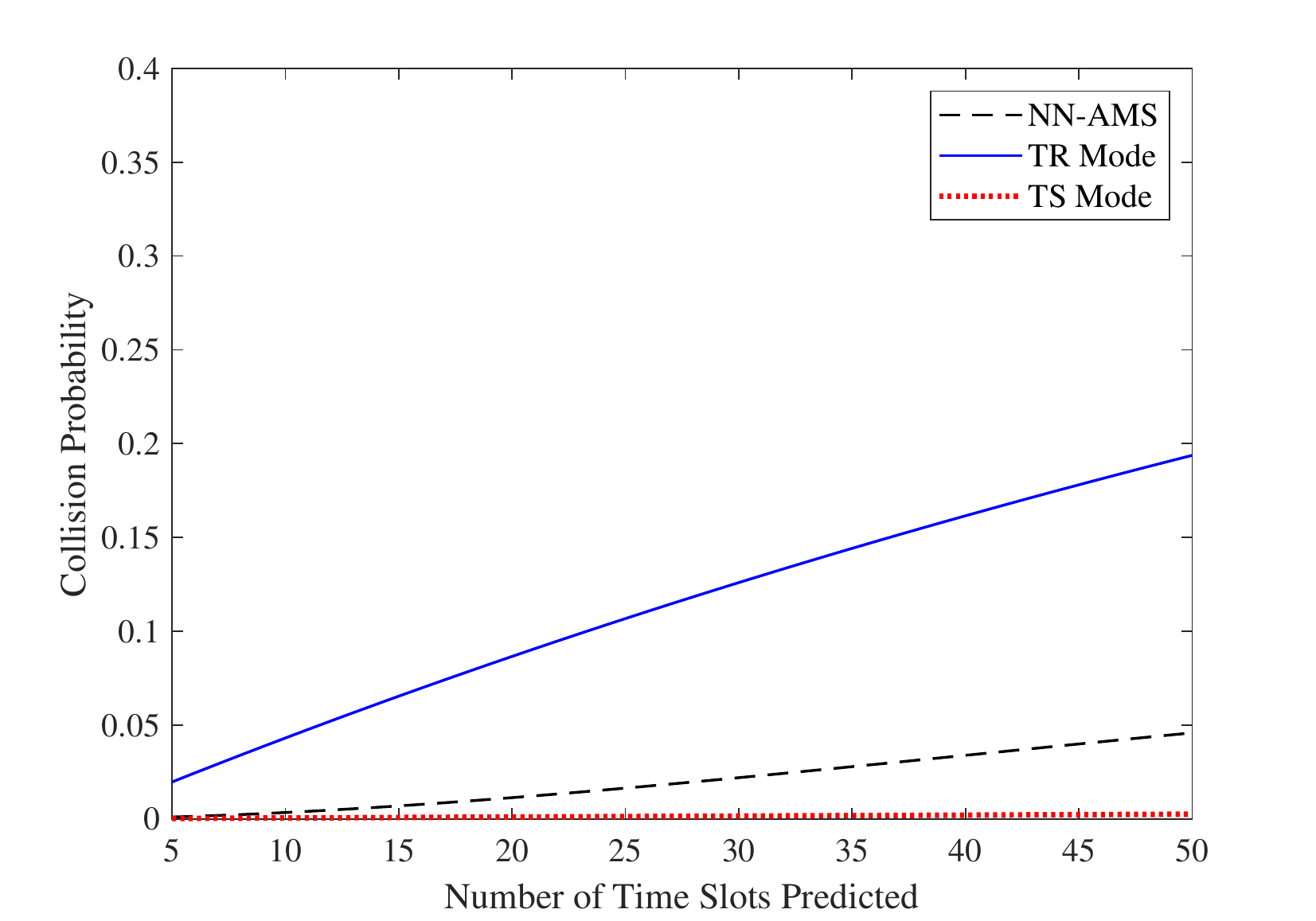} 
\caption{Collision probability versus the number of time slots \(M\).}
\label{Figure_6} 
\end{figure} 

\subsection{Collision Probability Analysis}

Figure. \ref{Figure_6} shows the collision probability versus number of time slots predicted $M$. The collision probabilities of all three schemes increase monotonically with the number of transmission slots. The reason is that the probability of PU's return to the channel increases with the length of transmission duration. The collision probability of TS mode is extremely low (i.e., close to zero) in practice. Thanks to the adaptive switch between TR and TS mode, the NN-AMS scheme gains much lower collision probability than that of TR mode. When \(M\) is small, the collision probability of NN-AMS scheme is very close to that of TS mode due to the low prediction error probabilities. It can be seen that the collision probability of NN-AMS scheme is only 8\% of that of TR mode when \(M=10\). Through jointly analysing the above figures, we conclude that our proposed NN-AMS scheme can dramatically reduce the collision probability at the expense of slightly lower SU's average throughput than that of conventional TR mode.

\section{Conclusion}
In this paper, we proposed a new application of NN in FD-CRNs. A multi-layer NN predictor and an AMS scheme were designed. Through testing the prediction performance, we analysed the influence of sensing error on the prediction error. We also derived SU's average throughput and collision probability of our proposed NN-AMS scheme. Simulation results show that SU can achieve almost the same average throughput in NN-AMS scheme as that of TR mode when PU is less tolerant of collision. Meanwhile, our proposed NN-AMS scheme reduces the collision probability by up to 92\% compared with conventional TR mode.

\bibliographystyle{IEEEtran}
\bibliography{NNAMS.bib}
\end{document}